\documentclass[sigconf, nonacm]{acmart}

\usepackage{tikz}

\newcommand{\sparagraph}[1]{\vspace{1mm}\noindent {\bf #1}}
\newcommand{\LIKE}{\texttt{LIKE}}
\newcommand{\REGEX}{\texttt{REGEX}}

\newcommand\vldbyear{2025}
\newcommand\vldbworkshop{Applied AI for Database Systems and Applications (AIDB 2025)}
\newcommand\vldbauthors{\authors}
\newcommand\vldbtitle{\shorttitle}
\newcommand\vldbavailabilityurl{https://github.com/utndatasystems/string-fingerprints}
\newcommand\vldbpagestyle{plain}

\usepackage{todonotes}

\usepackage{makecell}
\usepackage{siunitx}
\newcommand{\st}{\text{s.t.}}

\makeatletter
\newcommand{\fcdot}{\,\cdot\,}
\newcommand{\fcarg}[1]{\def\fc@rg{#1}\ifx\fc@rg\empty\fcdot\else\fc@rg\fi}
\makeatother

\newcommand{\abs}[1]{\lvert\fcarg{#1}\rvert}

\newcommand{\set}[1]{\{#1\}}

\usepackage{listings}
\definecolor{dkgreen}{rgb}{0,0.6,0}
\definecolor{gray}{rgb}{0.5,0.5,0.5}
\definecolor{mauve}{rgb}{0.58,0,0.82}
\definecolor{mygray}{rgb}{0.5, 0.5, 0.5}
\begin{document}
\title{Instance-Optimized String Fingerprints (Extended Abstracts)}

\author{Mihail Stoian}
\orcid{0000-0000-0000-0000} 
\authornote{The author contributed equally to this work.}
\affiliation{
  \institution{University of Technology Nuremberg}
  \city{Nuremberg}
  \country{Germany}
}
\email{mihail.stoian@utn.de}

\author{Johannes Th\"urauf}
\orcid{0000-0001-8516-6250}
\authornotemark[1]
\affiliation{
  \institution{University of Technology Nuremberg}
  \city{Nuremberg}
  \country{Germany}
}
\email{johannes.thuerauf@utn.de}

\author{Andreas Zimmerer}
\orcid{0000-0002-4158-5805}
\affiliation{%
  \institution{University of Technology Nuremberg}
  \streetaddress{}
  \city{Nuremberg}
  \state{Germany}
  \postcode{}
}
\email{andreas.zimmerer@utn.de}

\author{Alexander van Renen}
\orcid{0000-0002-6365-4592}
\affiliation{%
  \institution{University of Technology Nuremberg}
  \streetaddress{}
  \city{Nuremberg}
  \state{Germany}
  \postcode{}
}
\email{alexander.van.renen@utn.de}

\author{Andreas Kipf}
\orcid{0000-0003-3463-0564}
\affiliation{%
  \institution{University of Technology Nuremberg}
  \streetaddress{}
  \city{Nuremberg}
  \state{Germany}
  \postcode{}
}
\email{andreas.kipf@utn.de}

\begin{abstract}
Recent research found that cloud data warehouses are text-heavy. However, their capabilities for efficiently processing string columns remain limited, relying primarily on techniques like dictionary encoding and prefix-based partition pruning.

In recent work, we introduced string fingerprints---a lightweight secondary index structure designed to approximate \LIKE{} predicates, albeit with false positives. This approach is particularly compelling for columnar query engines, where fingerprints can help reduce both compute and I/O overhead. We show that string fingerprints can be optimized for specific workloads using mixed-integer optimization, and that they can generalize to unseen table predicates. On an IMDb column evaluated in DuckDB v1.3, this yields table-scan speedups of up to 1.36$\times$.
\end{abstract}

\maketitle

\pagestyle{\vldbpagestyle}
\begingroup\small\noindent\raggedright\textbf{VLDB Workshop Reference Format:}\\
\vldbauthors. \vldbtitle. VLDB \vldbyear\ Workshop: \vldbworkshop.\\ 
\endgroup
\begingroup
\renewcommand\thefootnote{}\footnote{\noindent
This work is licensed under the Creative Commons BY-NC-ND 4.0 International License. Visit \url{https://creativecommons.org/licenses/by-nc-nd/4.0/} to view a copy of this license. For any use beyond those covered by this license, obtain permission by emailing \href{mailto:info@vldb.org}{info@vldb.org}. Copyright is held by the owner/author(s). Publication rights licensed to the VLDB Endowment. \\
\raggedright Proceedings of the VLDB Endowment. 
ISSN 2150-8097. \\
}\addtocounter{footnote}{-1}\endgroup

\ifdefempty{\vldbavailabilityurl}{}{
\vspace{.3cm}
\begingroup\small\noindent\raggedright\textbf{VLDB Workshop Artifact Availability:}\\
The source code, data, and/or other artifacts have been made available at \url{\vldbavailabilityurl}.
\endgroup
}

\section{Introduction}\label{sec:introduction}

The last three years have shown us that natural language, and implicitly unstructured text, can be \emph{the} language we use to talk to machines~\cite{llama, instruct-gpt}. When it comes to the data itself, we can see a similar trend. Indeed, recent research shows that cloud data warehouses are rather text-heavy~\cite{DBLP:conf/sigmod/VogelsgesangHFK18, CAB, redset}. Yet, how advanced are our techniques to deal with such kind of data? Our work scratches the surface by proposing a lightweight secondary index that boosts queries, i.e., \LIKE{} (and simple \REGEX{}) predicates, on columns of this data type.

\begin{figure}
    \centering
    \includegraphics[width=1.0\linewidth]{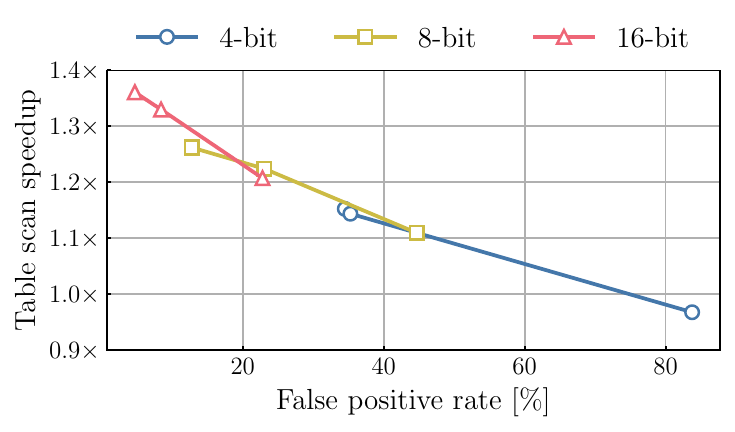}
    \caption{Using string fingerprints to speedup table scans with \LIKE{} predicates on IMDb's \texttt{title} column in DuckDB v1.3. Various instance-optimized partitions incur different false positive rates which correlate with the speedup achieved.}
    \label{fig:main-plot}
\end{figure}

\sparagraph{Motivation.} Transforming a data column to a sparser representation, which can still be queried---with possible allowance of false positives---enables many performance improvements; think of bloom filters~\cite{bloom-filter} and range filters~\cite{surf, infini-filter, memento-filter}, just to name a few from our research field. However, with the exception of a few examples, e.g., compression~\cite{fsst}, indexing~\cite{surf, n-gram-1st, n-gram-lp, regex-exp}, cardinality estimation~\cite{like-card-est-sigmod}, it seems that strings are still waiting for more attention, despite their prevalence as data type. Indeed, even recent progress on (updatable) bitmap indices focus on numeric data only~\cite{cubit}.

\sparagraph{String Fingerprints.} As a by-product of recent work on robust query processing~\cite{parachute}, we introduced string fingerprints as a mechanism that can mimic a \LIKE{} predicate, albeit with false positives. The key idea is to compactly represent the set of constituting letters of a string $S$ in a fixed number of bins that can be represented as a binary mask of fixed bitwidth (\#bins $\triangleq$ bitwidth). For a given pattern $P$, the generic form of a \LIKE{} predicate, $S.\texttt{contains}(P)$, evaluates to false if the fingerprint of $P$ is not a \emph{subset} of that of $S$.

\sparagraph{Applications.} This rather concise representation has a twofold role: (a) By attaching a fingerprint-column to the table, we can skip non-qualifying rows, and (b) if we maintain a dictionary with the fingerprint-values seen in the partition, we can use them to skip non-qualifying partitions. For instance, production systems offer the latter for prefix/suffix-based predicates only~\cite{snowflake}. In what follows, we focus on the former.

\sparagraph{Contribution.} Inspired by the recent interest of production systems towards instance-optimized database components~\cite{pred_cache, mddl}, we show that string fingerprints can be instance-optimized with respect to the workload and the data itself by using state-of-the-art mixed-integer optimization. In particular, on IMDb's \texttt{title} column of the JOB benchmark~\cite{job}, one can speed up table scans by up to 1.36$\times$ when using 16-bit fingerprints, as shown in Figure~\ref{fig:main-plot}. Moreover, we empirically show that the so instance-optimized fingerprints even generalize to \emph{unseen} table predicates.

\sparagraph{Related Work.} Work on indexing text using $n$-grams dates back to the seminal work of Ogawa and Matsuda~\cite{n-gram-1st}, and has been further developed---especially in the context of indexing for regular expressions~\cite{n-gram-2nd, best, free, n-gram-lp, regex-exp}---where the key idea is to select a near-optimal subset of $n$-grams to index (under constraints such as space). With string fingerprints, we argue that, particularly for individual letters ($1$-grams), there is no need to retain only a subset. Instead, we can index all grams to minimize the false positive rate.

\sparagraph{Overview.} We first introduce the preliminaries of string fingerprints and then present our main technical contribution, an approach to instance-optimize them using mixed-integer optimization (Sec.~\ref{subsec:mip}). We evaluate the approach on both seen and unseen predicates in Sec.~\ref{sec:evaluation}. We conclude with future work in Sec.~\ref{sec:conclusion}.

\section{Instance-Optimized Fingerprints}\label{sec:nussella}

\sparagraph{Preliminaries.} To better understand the intuition behind the MIP formulation in Sec.~\ref{subsec:mip}, let us next outline how fingerprints work.

\begin{figure}
    \centering
    \includegraphics[width=0.75\linewidth]{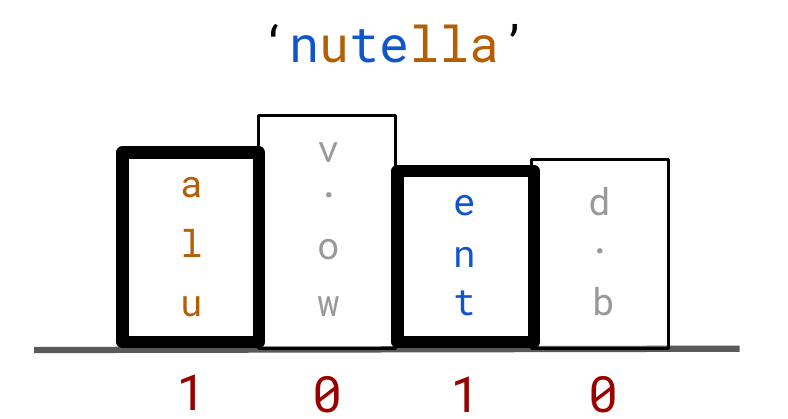}
    \caption{String fingerprints are bitmasks indexing letter bins}
    \label{fig:nutella}
\end{figure}

The main intuition is that a pattern $P$ has a chance to be contained in a string value $S$, i.e., $S.\texttt{contains}(P)$, if the letters of $P$ are a \emph{subset} of that of $S$. To see this, consider the pattern \texttt{'\%utn\%'} and the strings \texttt{'nutella'} and \texttt{'tone'}.
When computing their letter sets, we observe that the first \emph{could} qualify for the pattern, while the latter cannot, since the letter set of \texttt{'utn'} is \emph{not} a subset of \{\texttt{'t'}, \texttt{'o'}, \texttt{'n'}, \texttt{'e'}\}. On the other hand, even though the first string value qualifies, it is indeed a false positive. In particular, note that this representation cannot generate false negatives.

\sparagraph{Example.} Instead of indexing all letters, we first partition them. Consider the example in Figure~\ref{fig:nutella} for the string \texttt{'nutella'}. The letter space is partitioned in 4 bins (\#bins $\triangleq$ bitwidth). The letters \texttt{'a'}, \texttt{'l'}, and \texttt{'u'} fall into the first bin (read from left), while the others in the third bin. Hence, its fingerprint reads \texttt{1010}.
Similarly, we can compute the fingerprints of \texttt{'utn'} and \texttt{'tone'}, which are \texttt{1010} and \texttt{0110}, respectively---observe that the letter \texttt{'o'} falls into the second bin. In particular, note that the fingerprint of \texttt{'utn'} is indeed a subset of that of \texttt{'nutella'}, i.e., \texttt{1010} $\subseteq$ \texttt{1010}, while the same does not hold for \texttt{'tone'}, i.e., \texttt{0110} $\not\subseteq$ \texttt{1010}. Consequently, the partitioning produced a false positive for \texttt{'nutella'} and correctly classified \texttt{'tone'} as a true negative.

\subsection{Optimal Partitioning}\label{subsec:mip}

Mixed-integer linear optimization~\cite{Clautiaux_Ljubic:2025,
  wolsey2020integer, Conforti_et_al:2014} is a fundamental tool in operations
research,
which has been successfully applied in different domains
including supply chain management~\cite{Clautiaux_Ljubic:2025} and
machine learning~\cite{Bertimas_et_al:2019}.
Over the last decades enormous computational progress has been made by
developing new solution methods and enhancing existing ones leading to efficient
optimization solvers such as \textsf{Gurobi}~\cite{gurobi},
\textsf{CPLEX}~\cite{cplex}, and \textsf{SCIP}~\cite{scip}.
Combining the algorithmic advancements with modern
computational resources makes it possible to solve optimization problems that
were out of scope decades ago.

We now present the developed mixed-integer linear optimization model solving which yields an optimal partitioning.
Here, optimal means that for the given query patterns $\mathcal{Q}$ and a column's string-values---henceforth, words $\mathcal{W}$---the computed partitioning
maximizes the number of correctly classified data, which is equivalent to
minimizing the false positive rate.
In particular, if the optimization model is solved to global optimality, the
used mathematical methods guarantee that the resulting partitioning
has the lowest possible false positive rate for the given queries and data.

\sparagraph{Notation.} We introduce the necessary notation in Table~\ref{tab:notation}.
\begin{table}
  \centering
    \caption{Notation of the parameters used in the mixed-integer optimization
      model~(\ref{model-false-positive})}
    \label{tab:notation}
    \begin{tabular}{ll}
    Parameter & Description \\
    \toprule
    $\mathcal{A}$  & Set of characters to be partitioned
                     into the bins \\
    $n$ & Number of bins ($\triangleq$ bitwidth) \\
    $\mathcal{W}$ & Words, i.e., set of the column's string values \\
    $\mathcal{Q}$ & Set of given query patterns \\
    $f(\cdot)$ & \makecell[l]{Function that returns for a query the set of words \\ in $\mathcal{W}$ that contain the query as substring} \\
    $s_{i}$  & Denoting the $i$th character of string~$s$ \\
    $len(s)$ & Denoting the length of string~$s$ \\
    $[l]$ & For~$l \in \mathbb{N}$, it represents the set $\set{1, \ldots,
            l}$ \\
    \bottomrule
  \end{tabular}
\end{table}
In addition, we use the following optimization variables.
For each bin~$i \in [n]$ and character~$a \in \mathcal{A}$, the binary
variable~$x_{a,i} \in \set{0,1}$ evaluates to~$1$ if character~$a$ is
in bin~$i$. Otherwise, $x_{a,i} = 0$ holds.
Thus, the $x$ variables determine the optimized partitioning of the characters into
the bins.
For each string~$s \in \mathcal{Q} \cup \mathcal{W}$ and bin~$j$, the binary
variable~$d^{s}_{j} \in \set{0,1}$ indicates if there is a letter of
string~$s$ that is in bin~$j$, i.e., $d^{s}_{j}=1$ holds, and if this is
not the case, $d^{s}_{j} = 0$ is satisfied.
Consequently, $d^{s}_{j}, j \in [n]$, represent the string fingerprint of
string~$s$.
For each query~$q \in \mathcal{Q}$ and word~$w \in \mathcal{W}
\setminus f(q)$, the
variable~$\eta^{w,q} \in \set{0,1}$ evaluates to one if the optimized partitioning
correctly determines that query~$q$ is not
contained in word~$w$, i.e., the optimized partitioning does not produce a false
positive for the considered query~$q$ and word~$w$.
However, if the optimized partitioning produces a false positive for this query and
word, then $\eta^{w,q} = 0$ holds.

\sparagraph{Model.} The aim of the optimization model is to maximize the number of
correctly classified query-word combinations, i.e., to maximize~$\sum_{q \in
  \mathcal{Q}} \sum_{w \in \mathcal{W} \setminus f(q)}\eta^{w,q}$.
We note that this is equivalent to minimizing the number of falsely classified
query-string combinations.
Using the introduced variables and notation, the optimization model to
compute an optimal partitioning w.r.t. the given query-word combinations is given
by
\begin{subequations} \label{model-false-positive}
  \begin{align}
    \max_{x, d, \eta} \quad & \sum_{q \in \mathcal{Q}} \sum_{w \in \mathcal{W} \setminus f(q)}
                              \eta^{w,q} \label{model-objective-function}
    \\
    \st \quad
    & \sum_{j=1}^{n} x_{a,j} = 1, \quad a \in
      \mathcal{A}, \label{cons:partition-letters-bin} \\
    & x_{s_{i},j} \leq  d_{j}^{s}, \quad i \in [\text{len}(s)], \ j \in [n], \
      s \in \mathcal{Q} \cup \mathcal{W}, \label{cons:determine-ids--part1} \\
    & d^{s}_{j} \leq \sum_{i \in [\text{len}(s)]} x_{s_{i},
      j} \label{cons:determine-ids-part2}, \quad j \in [n], \ s \in \mathcal{Q}
      \cup \mathcal{W}, \\
     & d^{q}_{j} \leq d^{w}_{j}, \quad j \in [n], \ q \in \mathcal{Q}, \ w \in
       f(q), \label{cons:no-false-negative} \\
    & \eta^{w,q} \leq \sum_{j=1}^{n} (1-d^{w}_{j})d^{q}_{j}, \quad w \in
      \mathcal{W} \setminus f(q), \ q
      \in \mathcal{Q}, \label{cons:determin-correct-strong-wo-query} \\
    &
    \begin{aligned}
    & x \in \set{0,1}^{\abs{\mathcal{A}} \times n}, \ d \in \set{0,1}^{(\abs{\mathcal{W}} +
      \abs{\mathcal{Q}}) \times n}, \\
    & \eta \in \set{0,1}^{\abs{\mathcal{W}}
      \times \abs{\mathcal{Q}} - \sum_{q \in \mathcal{Q}} \abs{f(q)}}.
    \end{aligned} \label{model:variable-definitions}
\end{align}
\end{subequations}
\sparagraph{Model Description.}
In Constraints~\eqref{cons:partition-letters-bin}, we ensure that each character of~$\mathcal{A}$ is exactly assigned to one bin.
By Constraints~\eqref{cons:determine-ids--part1}
and~\eqref{cons:determine-ids-part2}, for each string~$s \in
\mathcal{Q} \cup \mathcal{W}$ we determine
the string fingerprint~$d^{s}$ w.r.t.\ the chosen partitioning given by~$x$.
More precisely, $d^{s}_{j}$ equals one if and only if string~$s$ contains at
least one letter that is in bin~$j$. Otherwise, it is zero.
Constraints~\eqref{cons:no-false-negative} ensure that we do not have any false
negatives by enforcing that every string that contains the query has a string
fingerprint that includes the one of the query.
We now consider a query~$q \in \mathcal{Q}$ and word~$w \in \mathcal{W}$ that does not
contain query~$q$, i.e., $w \in \mathcal{W} \setminus f(q)$.
If $d^{q}_{j} \leq d_{j}^{w}$ for all $j \in [n]$ holds, then we have wrongly
classified the word~$w$ to contain $q$.
Further, if $d^{q}_{j} \leq d_{j}^{w}$  is satisfied for all $j \in [n]$,
the right-hand side of Constraint~\eqref{cons:determin-correct-strong-wo-query}
evaluates to zero due to $d^{q}_{j}, d_{j}^{w} \in \set{0,1}$.
Consequently, in this case
Constraint~\eqref{cons:determin-correct-strong-wo-query} implies $\eta^{w,q} =0$
and we cannot increase the objective function by wrongly classified strings.
However, if $d^{q}_{j} \leq d_{j}^{w}$ does not hold, i.e., we correctly
determine that word~$w$ does not contain~$q$, then the right-hand side of
Constraint~\eqref{cons:determin-correct-strong-wo-query} is at least one.
Consequently, we can set $\eta^{w,q} = 1$ and increase the objective function by
correctly identifying that word~$w$ does not contain $q$.
The objective function~\eqref{model-objective-function} then maximize the number of
correctly classified strings that do not contain the corresponding query, which is
equivalent to minimizing the number of false positives.
Consequently, solving the presented optimization model to global optimality
leads to an optimized partitioning that minimizes the false positive rate
w.r.t.~the given queries and data.

We note that Model~(\ref{model-false-positive}) is a mixed-integer nonlinear
optimization problem in the presented form due to the products of binary
variables in
Constraints~(\ref{cons:determin-correct-strong-wo-query}).
However, using standard techniques of mixed-integer optimization, these products
of binary variables can be equivalently reformulated with the help of additional
variables and linear constraints, e.g.,  see~\cite{Glover_Woolsey:1974}.
This leads to a mixed-integer linear optimization model, which we used in our
computational study.
We further neglect Constraints~(\ref{cons:no-false-negative}) in our
computations because they are mathematically redundant, i.e., they can be
removed without changing the set of optimal solutions.
This directly follows from the fact that for $d_{j}^{q} = 1$ there exists a
character in bin~$j$ that is included in query~$q$ due to
Constraints~(\ref{cons:determine-ids--part1})
and~\eqref{cons:determine-ids-part2}.
However, this character is also included in word~$w$ due to $q \in \mathcal{Q}$
and $w \in f(q)$.
Consequently, Constraints~\eqref{cons:determine-ids--part1} implies $d_{j}^{w}=
1$ and then the corresponding Constraint~(\ref{cons:no-false-negative}) is
satisfied. The latter is directly valid for the case $d_{j}^{q} = 0$.
This formal discussion proves
that the computed partitioning and corresponding string fingerprints cannot produce
false negatives.

Solving Model~(\ref{model-false-positive}) to global optimality using
state-of-the-art solvers computes a partitioning that is optimal w.r.t. the
considered queries and data, i.e., it has the lowest possible false positive
rate.
However, for a large number of queries, words, and bins, the number of
variables and constraints is enormous, which makes it challenging to solve
these models to global optimality in a reasonable time.
To ensure that the computational time stays within practical limits, a time limit
can be imposed on the optimization process.
As shown in the evaluation, this leads to computing a high-quality
partitioning, that may not be optimal, but efficiently
minimizes the false positive rate within the time limit.
Moreover, the following computational results show that the optimized partitioning
also performs well for unseen queries and data.

\begin{figure*}
    \centering
    \begin{tikzpicture}
        \node[anchor=south west,inner sep=0] (image) at (0,0) {\includegraphics[width=1.0\linewidth]{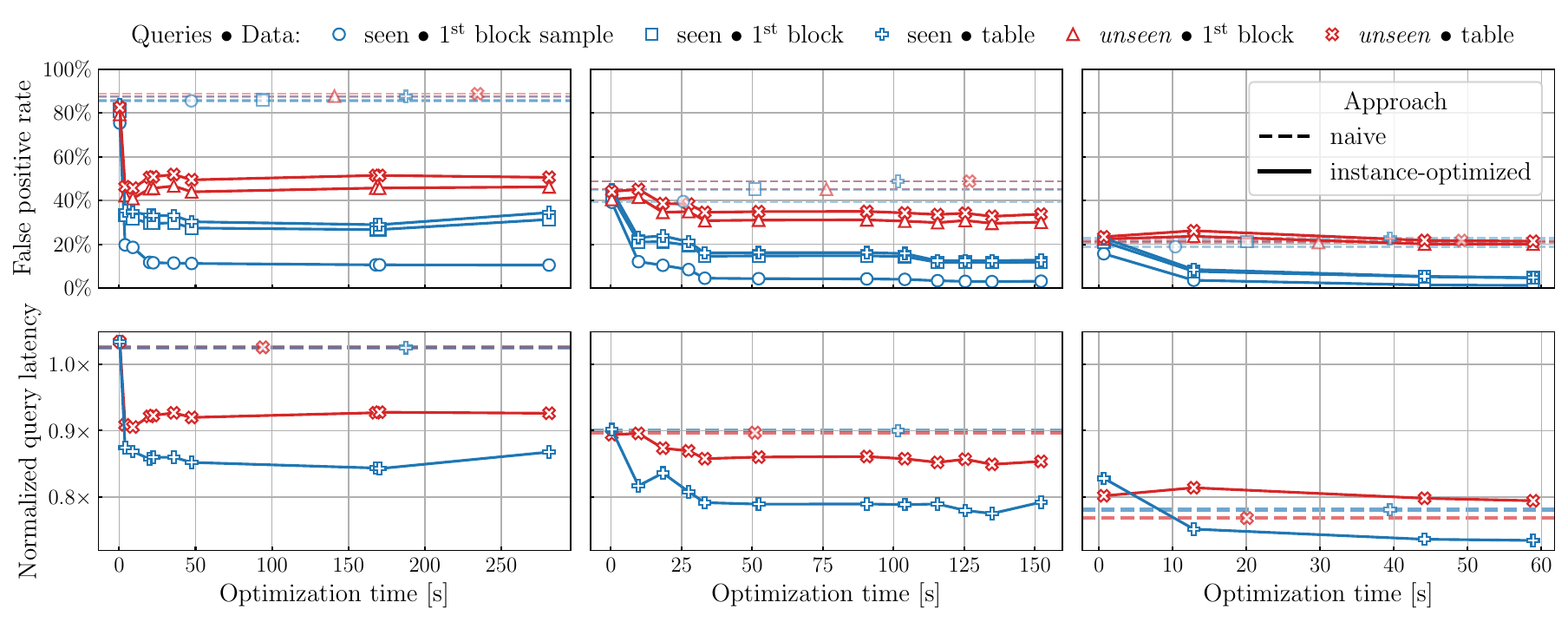}};
        \begin{scope}[x={(image.south east)},y={(image.north west)}]
            \node at (0.22,-0.05) {\textbf{(a) 4-bit}};
            \node at (0.525,-0.05) {\textbf{(b) 8-bit}};
            \node at (0.85,-0.05) {\textbf{(c) 16-bit}};
        \end{scope}
    \end{tikzpicture}
    \caption{Effect of string fingerprints on table scans over IMDb’s \texttt{title} column for various fingerprint bitwidths $\in \{4, 8, 16\}$ and (query, data) pairs. The 300-query workload consists of the 10 highest-, mid-, and lowest-frequency column's $k$-grams for each $k \in \{1, \ldots, 10\}$. Queries are split into 20 \emph{seen} and 280 \emph{unseen} patterns. Instance-optimized partitions are trained on a 50-tuple sample from the first data block using the seen queries. We also plot a subset of the intermediate solutions obtained by the solver during the optimization process. The naive, workload-agnostic baseline assigns characters in a round-robin manner.}
    \label{fig:full-plot}
\end{figure*}
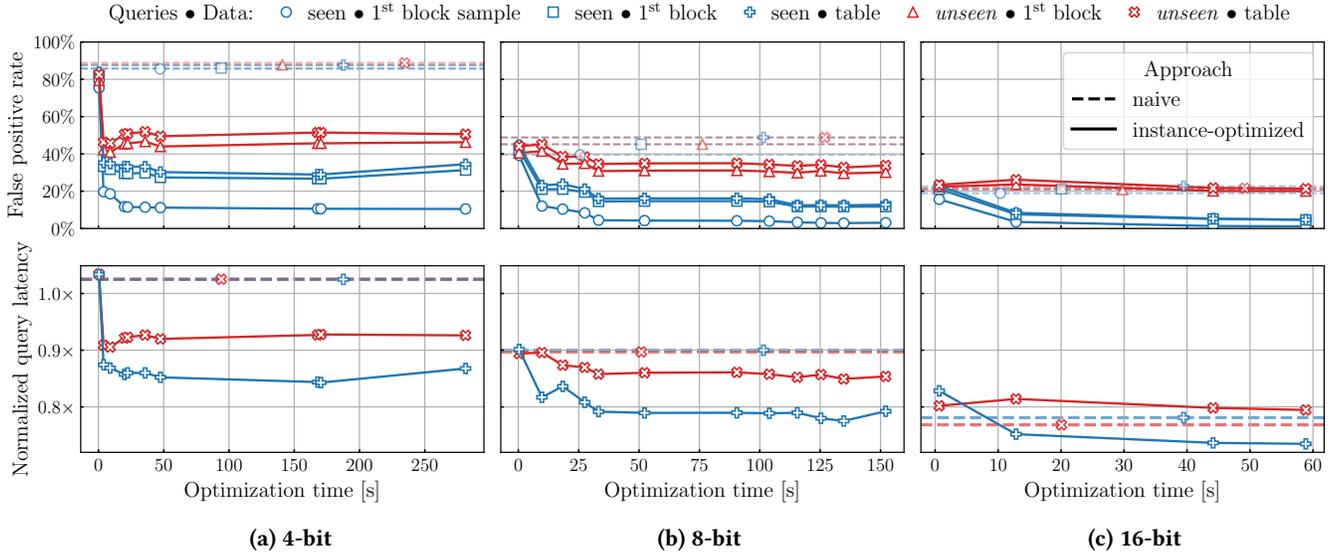
\section{Evaluation}\label{sec:evaluation}

\sparagraph{Setup.}~We run the queries single-threaded on a single node
Intel\textsuperscript{\textregistered} Xeon\textsuperscript{\textregistered}
Gold 5318Y CPU (24 cores, 48 hyper-threads). The machine is equipped with 128GB
DDR4 main memory and runs Ubuntu 24.04. We use DuckDB v1.3.0 as query engine. The MIP solutions are computed with \textsf{Gurobi}~12.0.1 (\cite{gurobi}), with a time limit of~\SI{300}{\second} and a thread limit of 48. In the following, we denote as optimization time the total time required for reading the data, building the optimization model, and solving it.

\sparagraph{Benchmark.} We consider the real-world IMDb dataset from the JOB benchmark~\cite{job} and take as reference its \texttt{title} column with \SI{2.53}{\text{M}} movie names; since we currently optimize for printable bytes only (100 distinct ones, i.e., $|\mathcal{A}|$ = 100), the table is reduced to \SI{2.37}{\text{M}} tuples. We compose a 300-query workload consisting of the 10 highest-, mid-, and lowest-frequency $k$-grams for each $k \in \{1, \ldots, 10\}$ extracted from the column. These are randomly split into 20 seen and 280 unseen queries.

Due to DuckDB v1.3.0's limited support for complex predicate pushdown, we simulate the pushdown of our bitmask check by (a) measuring the time to perform the bitmasked table scan, (b) building an auxiliary column that indicates the result, and (c) and measuring the time of the new query. The reported time is (a) + (c).

\subsection{Table Scan Speedup}

One of the main applications of string fingerprints is accelerating table scans. Namely, in the context of a columnar query engine, one can attach the fingerprint column and, at query time, evaluate the corresponding predicate first:
\begin{lstlisting}[label=l:nutella-example,language=SQL]
where [...] title_fp & pattern_mask = pattern_mask
and title like '%(*@\textcolor{mygray}{\{pattern\}}@*)%' [...].
\end{lstlisting}
In this case, the \texttt{LIKE} predicate may need to be evaluated on significantly fewer tuples. This effect is visualized in Figure~\ref{fig:full-plot}, where, we show the false positive rates (FPRs) for various fingerprints bitwidths $\in \{4, 8, 16\}$ (upper subplots) and the query latencies for seen and unseen queries on the full table (lower subplots).

\sparagraph{$1^{\mathrm{st}}$ Observation.} Despite being trained on a random 50-tuple sample of the first data block (=$2^{16}$ tuples) of the table, we observe that the FPRs of the instance-optimized fingerprints remain competitive even on the full table. This is also reflected in the latency numbers, where in the 16-bit setting, the speedup reaches 1.36$\times$. The same holds for the unseen queries, for which the attained speedup reads 1.26$\times$. The optimization time spent for this setting amortizes for the unseen queries already in the 4th run of the workload.

\sparagraph{$2^{\mathrm{nd}}$ Observation.} Given the fact that the pattern lengths are bounded above by 10, the instance-optimized setting is not worth it with increasing bitwidth. This is due to the fact that with a larger bitwidth (and a rather bounded alphabet), the bin densities are much lower, thus allowing for sparser fingerprints, both for patterns and column string values.

\sparagraph{$3^{\mathrm{nd}}$ Observation.}
For the 4-bit and 8-bit setting, the underlying optimization problems to compute
the partitions in Figure~\ref{fig:full-plot} cannot be solved to global optimality
within \SI{300}{\second}.
However, we obtain a MIP optimality gap of
around 5.6\% and 1.9\%, respectively. Consequently, the
quality of the current best partition is rather close to the one of an optimal
solution (w.r.t.~the given query-data). In particular, optimizing for at most \SI{60}{\second} is sufficient to produce solutions that are competitive with the fastest in terms of query performance; for the 8-bit case, the gap reads 1.82\%.
\section{Conclusion \& Future Work}\label{sec:conclusion}

String fingerprints are a promising research direction towards efficiently querying string columns. The key idea is to partition the letter space such that the false positive rate on a given workload is minimized. More importantly, unlike recent caching mechanisms~\cite{pred_cache, mddl}, string fingerprints (a) can be both used to instance-optimize the database and (b) generalize to \emph{unseen} queries. Regardless of the extent of predicate pushdown in the system, instance-optimized string fingerprints act as table scan accelerators by reducing the number of actual string predicate evaluations.

\sparagraph{Fingerprints $\Join$ $N$-grams.} To achieve even better FPRs, one could consider using larger grams instead of letters (1-grams) only. The presented MIP formulation (Sec.~\ref{subsec:mip}) can be applied to this setting with adaptations, however, the corresponding optimization problems become more challenging since the alphabet ($\mathcal{A}$; Tab.~\ref{tab:notation}), and hence the number of optimization variables, naturally increase. Using larger grams fits the line of research on $n$-gram indexing for regular expressions~\cite{n-gram-1st, n-gram-2nd, best, free, n-gram-lp, regex-exp}, which instead optimizes which subset of $n$-grams to index.

In future work, we also plan to investigate how string fingerprints can be used for (a) string zonemaps---particularly for large bitwidths, (b) \LIKE{} cardinality estimation~\cite{string-card-est-trad1, astrid-card-est, string-card-est-trad2, lplm-card-est, like-card-est-sigmod}, and (c) clustering, even across multiple string columns.

\sparagraph{Acknowledgments.} The authors gratefully acknowledge the scientific support and HPC resources provided by the Erlangen National High Performance Computing Center (NHR@FAU) of the Friedrich-Alexander-Universität Erlangen-Nürnberg (FAU). The hardware is funded by the German Research Foundation (DFG).

The authors acknowledge the use of DeepL and OpenAI's ChatGPT for partly editing and polishing the text and figures for spelling, grammar, and stylistic improvements. Additionally, ChatGPT was utilized for support in basic coding tasks.

\balance


\bibliographystyle{ACM-Reference-Format}
\bibliography{sample}

\end{document}